\documentclass[aps,prd,twocolumn,article]{revtex4}
\usepackage{ulem}
\usepackage{color}
\usepackage{graphicx}
\usepackage{epsfig}

\usepackage{bm}
\newcommand{\be}{\begin{equation}}
\newcommand{\ee}{\end{equation}}
\newcommand{\ba}{\begin{eqnarray}}
\newcommand{\ea}{\end{eqnarray}}

\begin{document}
\title{Estimating the Charm Quark Diffusion Coefficient and thermalization time from D meson spectra at RHIC and LHC} 
\author{Francesco Scardina$^{a,b}$, Santosh K. Das$^{a}$}
\author{Vincenzo Minissale $^{a,b}$, Salvatore Plumari$^{a,b}$, Vincenzo Greco$^{a,b}$}

\affiliation{$^a$ Department of Physics and Astronomy, University of Catania, 
Via S. Sofia 64, 1-95125 Catania, Italy}
\affiliation{$^b$ Laboratori Nazionali del Sud, INFN-LNS, Via S. Sofia 62, I-95123 Catania, Italy}
\date{\today}

\begin{abstract}
We describe the propagation of charm quarks in the quark-gluon plasma (QGP) by means of a Boltzmann transport approach. 
Non-perturbative interaction between heavy quarks and light quarks  have been taken into account through a 
quasi-particle approach in which light partons are dressed with thermal masses tuned to lQCD thermodynamics. 
Such a model is able to describe the main 
feature of the non-perturbative dynamics: the enhancement of the interaction strength near $T_c$. 
We show that the resulting charm in-medium evolution is able to correctly predict simultaneously 
the nuclear suppression factor, $R_{AA}$,  and the elliptic flow, $v_2$, at both RHIC and LHC energies and at different
centralities. The hadronization of charm quarks is described by mean of an hybrid model of fragmentation plus coalescence
and plays a key role toward the agreeement with experimental data.

We also performed calculations within the Langevin approach which can lead to very similar $R_{AA}(p_T)$ as 
Boltzmann, but the charm drag coefficient as to be reduced by about a $30\%$ and also generates an elliptic flow $v_2(p_T)$ is about a $15\%$
smaller. We finally compare the space diffusion coefficient $2\pi TD_s$ extracted by our phenomenological approach to lattice QCD results,
finding a satisfying agreement within the present systematic uncertainties.
Our analysis implies a charm thermalization time, in the $p\rightarrow 0$ limit, of about $4-6 \, fm/c$ which is smaller than the 
QGP lifetime at LHC energy.

\vspace{2mm}

\end{abstract}
\maketitle

\section{Introduction}
The study of QCD matter under extreme conditions of 
high temperatures is the primary purpose of ultra-relativistic heavy ion collisions
which are being performed at the  Relativistic Heavy-ion Collider (RHIC) and at 
the Large Hadron Collider (LHC). The energy deposited during the collisions 
produce a medium consisting of deconfinated quarks and gluons called Quark Gluon Plasma
(QGP)\cite{Shuryak:2004cy,Science_Muller}. An essential role to characterize the QGP can be played by the hard probes 
created in the initial stage of the collisions. Among them heavy quarks (HQs), charm and bottom, provide
a very promising probe since they travel through the expanding medium interacting with the light particles but  their 
number is expected to be conserved due to the large $M/T$ ratio. Therefore HQ can probe the whole evolution 
of the QGP and produced out-of-equilibrium are expected to conserve memory of the history of the plasma evolution 
\cite{hfr,Andronic:2015wma,Prino:2016cni, Aarts:2016hap, Greco:2017qm}.

Moreover HQ production can be calculated in next to leading order pQCD scheme and before the first experimental results it was expected 
that their interaction with the medium could be characterized by means of perturbative QCD which led to the expectations 
of a small suppression of the spectra and a small elliptic flow. However the first observations  of non photonic electrons coming 
from heavy quark decays measured in Au+Au at  $\sqrt{s_{NN}}$ = 200 GeV at RHIC \cite{phenixelat,stare,phenixe} shown a surprisingly 
small $R_{AA}$ and a quite large elliptic flow $v_2$, indicating a quite strong interactions between HQ and the medium which 
is substantially beyond the expectations from perturbative QCD \cite{vanHees:2004gq,moore,rappv2}. 
These observations triggered many studies in which non perturbative 
approach have been implemented. One of this approach consists of including non-perturbative contributions~\cite{rappprl} from the 
quasi-hadronic bound state with a subsequent hadronization  by coalescence and fragmentation \cite{Greco:2003vf,Greco:2003xt}. 
Other approaches make use of a pQCD framework supplemented by Hard Thermal Loop (HTL) in order 
to evaluate Debye mass and running coupling constant ~\cite{gossiauxv2,alberico}.
Another efficient way is to use a quasi-particle approach in which non perturabative effects are considered  
by introducing a thermal mass for the particle in the bulk, $m(T) \sim g(T) T$. A fit to lQCD thermodynamics
allows to determine $g(T)$ \cite{Berrehrah:2014kba,salvo}. 
All these models are based on collsional energy-loss which should be the dominant mechanism 
in the low momentum region of charm spectra \cite{bass,bassnpa,Cao:2016gvr}, $p_T \lesssim 3-5 M_{HQ}$, while at higher momenta 
there is a consensus that radiative energy loss becomes dominant even if self-consistently collisional
energy loss can never be discarded \cite{Djordjevic:2011dd,Cao:2016gvr,Gossiaux:2010yx,Das:2010tj}. 
Furthermore in the high-$p_T$ region pQCD schemes have shown 
to be able to account for the observed suppression of the spectra \cite{Djordjevic:2011dd,Djordjevic:2013xoa,Kang:2016ofv}
and some group obtained also a satisfying prediction for the elliptic flow \cite{Cao:2016gvr,Uphoff:2014cba}.

In this present paper we will focus on the results of a quasi-particle model (QPM) for charm quarks.
In Ref. \cite{salvo} it is shown that the quasi particle approach is able to reproduce the lattice QCD equation of state.
The extracted coupling $g(T)$ appears to have a significant deviation from pQCD especially
a $T\rightarrow T_c$. This leads to  a weakly T dependent drag coefficient $\gamma(T)$ ~\cite{Das:2015ana} at variance with
pQCD with a constant coupling $g$ or AdS/CFT where both predict a $T^2$ dependence drag coefficients. 
Such a feature of QPM has been found by other groups \cite{Berrehrah:2014kba,Song:2015ykw} that has also shown that the pattern remains
quite similar even when quasi-particle widths (off-shell dynamics) are accounted for. It has been thoroughly studied 
in Ref. \cite{Das:2015ana} for heavy quarks and also in \cite{Liao:2008dk, Scardina:2010zz} for the light sector that an interaction increasing 
as the the temperature decreases is one of the key ingredient to generate a larger elliptic flow and thus reducing 
the tension between the $R_{AA}$ and $v_{2}$ observed experimentally and calculated theoretically. 
This along with an hadronization via colescence is also a main underlying reason of  the  early T-matrix approach 
applied at RHIC energy \cite{rappv2,rappprl} and the following developments in \cite{he1}.
In the present work we employ the quasi particle approach as discussed in ref. \cite{Das:2015ana}. 

The main difference in this work with respect to the results presented in \cite{Das:2015ana} is  
the framework used to describe the heavy quark propagation based on a Boltzmann transport as well as the bulk evolution. Furthermore 
while we already performed an analysis of such differences (in between Langevin and Boltzmann) for schematic cases
like box calculations \cite{fs}, here we present the results for AA collisions at both RHIC and LHC energies and different centralities.
Furthermore we also include here a fragmentation plus coalescence for heavy quark hadronization and
moreover we discuss the difference entailed in terms of the spatial diffusion coefficient $D_s$ 
and compare them to lattice QCD results. 

The article is organized as follows. In section II, we discuss briefly
the Boltzmann transport equation and the Quasi-Particle approach for HQs. In section III, we describe the hybrid model of 
fragmentation and coalescence to consider the hadronization process of heavy quarks into heavy flavor mesons in QGP. 
Section IV is devoted to the comparison between the simulation results with experimental results at different colliding energy
and different centralities. In section V, we discuss the heavy quark transport coefficient obtained within the present approach.
Section V contains a summary and some concluding remarks.

\section{Transport equation for charm quarks in the QGP}
The evolution of the charm quark distribution function is obtained solving the relativistic Boltzmann 
transport equations~\cite{Ferini:2008he, fs} for charm quarks scattering in a bulk medium of quarks and gluons: 
\ba
 p^{\mu} \partial_{\mu}f_{Q}(x,p)= {\cal C}[f_q,f_g,f_{Q}](x,p) \nonumber  \\
 p^{\mu}_q \partial_{\mu}f_{q}(x,p)= {\cal C}[f_q,f_g](x_q,p_q)  \nonumber  \\
 p^{\mu}_g \partial_{\mu}f_{g}(x,p)= {\cal C}[f_q,f_g](x_g,p_g)   
\label{B_E} 
\ea
where $f_k(x,p)$ is the on-shell phase space one-body distribution function for the $k$ parton and
${\cal{C}}[f_q, f_g, f_{Q}](x,p)$ is the relativistic Boltzmann-like collision integral and the phase-space
distribution function of the bulk medium consists of quark and gluons entering the equation for charm quarks as
an external quantities in ${\cal{C}}[f_q,f_g,f_{Q}]$. 
We assume that the evolution of $f_q$ and $f_g$ are independent of $f_{Q}(x,p)$ and discard collisions between heavy quarks
which is by far a solid approximation.  
We are interested in the evolution of the HQ distribution function $f_{Q}(x,p)$. The evolution of the bulk of quark and gluons is 
instead given by the solution of the other two transport equations where the ${\cal C}[f_q,f_g]$ is tuned to a fixed $\eta/s(T)$, as discussed
in detail in ref.~\cite{Ruggieri:2013ova}.  This is however quite equivalent to a modeling where the bulk is given by
viscous hydrodynamics.

The collision integral for heavy quarks is given by
\begin{eqnarray}
{\cal C}[f_{Q}]&=&\frac{1}{2E_1}\int \frac{d^3p_2}{2E_2(2\pi)^3} \int\frac{d^3p_{1}^{\prime}}{2E_{1'}(2\pi)^3} \nonumber\\
&&\times \left[f_Q(p_{1}^{\prime})f_{q,g}(p_{2}^{\prime}) - f_Q(p_1) f_{q,g}( p_2)\right] \nonumber\\
&&\times |{\cal M}_{(q,g)+Q}(p_1p_2\rightarrow p_1^\prime p_2^\prime)|^2 \nonumber \\
&&  \times(2\pi)^4  \delta^4(p_1 + p_2 - p_{1}^{\prime} - p_{2}^{\prime})~,
\label{CollisionInt0}
\end{eqnarray}
where ${\cal M}_{{(q,g)+Q} \leftrightarrow {(q,g)+Q}}$ corresponds to the transition amplitude of the HQ scatterings.
In order to solve the collision integral it is necessary to evaluate the scattering matrix of the microscopical process.
In the present paper this is done in the framework of a Quasi-Particle model as described in the following.

The evolution of the QGP bulk given by 
an approach in which we  gauge the collision integral to the wanted $\eta/s$ as described in
 \cite{Plumari:2012ep,Ruggieri:2013bda,Ruggieri:2013ova,Plumari:2015cfa}.
In this way we are able to simulate the dynamical evolution of a fluid with specified $\eta/s$
by means of the Boltzmann equation. In the case considered here we have, more specifically
employed a bulk with massive quarks and gluons that  provide the possibility to have a softening of the
equation of state with a decreasing speed of sound when the
cross over region is approached. Within this approach we describe the
evolution of a system that dynamically has approximatively the lQCD
equation of state \cite{Borsanyi:2010cj}.
As shown in \cite{Plumari:2015sia} within this approach we recover universal
features of hydrodynamics and it permits to study the impact of
$\eta/s(T)$ on observables like $v_{n}(p_{T})$ in analogy to what is done within hydrodynamical 
simulations \cite{Romatschke:2007mq,Song:2011hk,Schenke:2010nt,Niemi:2011ix}.

The numerical solution of the Boltzmann equation is obtained by means of the test particle method
to map the one body distribution and we divide the space in a three-dimensional grid.
For being in a regime of convergency we employ a number of test particle per real particle
of 400, which we have verified allow to give a good convergency also for differential observables
like $v_2(p_T)$. More generally it has been checked that the numerical solution of the Boltzmann equation
for HQs leads to the Boltzmann-Juttner equilibrium distribution function in all the 
relevant momentum range.

The key role is certainly played by the scattering matrix ${\cal M}_{{(q,g)+Q} \leftrightarrow {(q,g)+Q}}$ 
that is the kernel of the interaction that allows also to calculate the drag and diffusion transport coefficients.
The ingredient of the Quasi Particle model are the 
thermal masses: $m^2_g(T)=3/4 g^2(T) T^2$, $m^2_{u,d}(T)=1/3 g^2(T) T^2$ and 
$m^2_{u,d}(T) - m^2_{0s} =1/3 g^2(T) T^2$.  The parametrized form of the strong coupling constant $g(T)$, 
is evaluated by making a fit of the energy density obtained by lattice QCD calculations 
and in our case has been parametrized as:
\begin{equation}
\label{Peshier_g_T}
g^2(T)=\frac{48 \pi^2}{(11 N_C-2 N_f)\ln{\left[\lambda(\frac{T}{T_C}-\frac{T_S}{T_C})\right]^2}}.
\end{equation}
where $N_c=N_f=3$, $\lambda=2.6$ and $T_s/T_c=0.57$.
It has been shown in ~\cite{salvo} that QPM is able to reproduce with good accuracy the lattice QCD pressure and interaction measure
$T^\mu_\mu=\epsilon-3 P$. The main feature of this approach is that the resulting coupling is significantly
stronger than the one coming from pQCD running coupling, particularly as $T\rightarrow T_c$.
The evaluation of the scattering matrix ${\cal M}_{{(q,g)+Q} \leftrightarrow {(q,g)+Q}}$  is then performed
considering the leading-order diagram with the effective coupling $g(T)$ that leads to effective vertices and a
dressed massive gluon propagator for $qQ \leftrightarrow qQ$ and massive quark propagator for $gQ\leftrightarrow gQ$
scatterings. The detail of the calculations for all $u,t,s$ channels and their interferences is quite long even if proceed along
a standard procedure and can be found in Ref. \cite{Berrehrah:2013mua}, where also a comparison with the massless case
and the massive including collisional widths is presented.

\begin{figure}[ht]
\begin{center}
\includegraphics[width=18pc,clip=true]{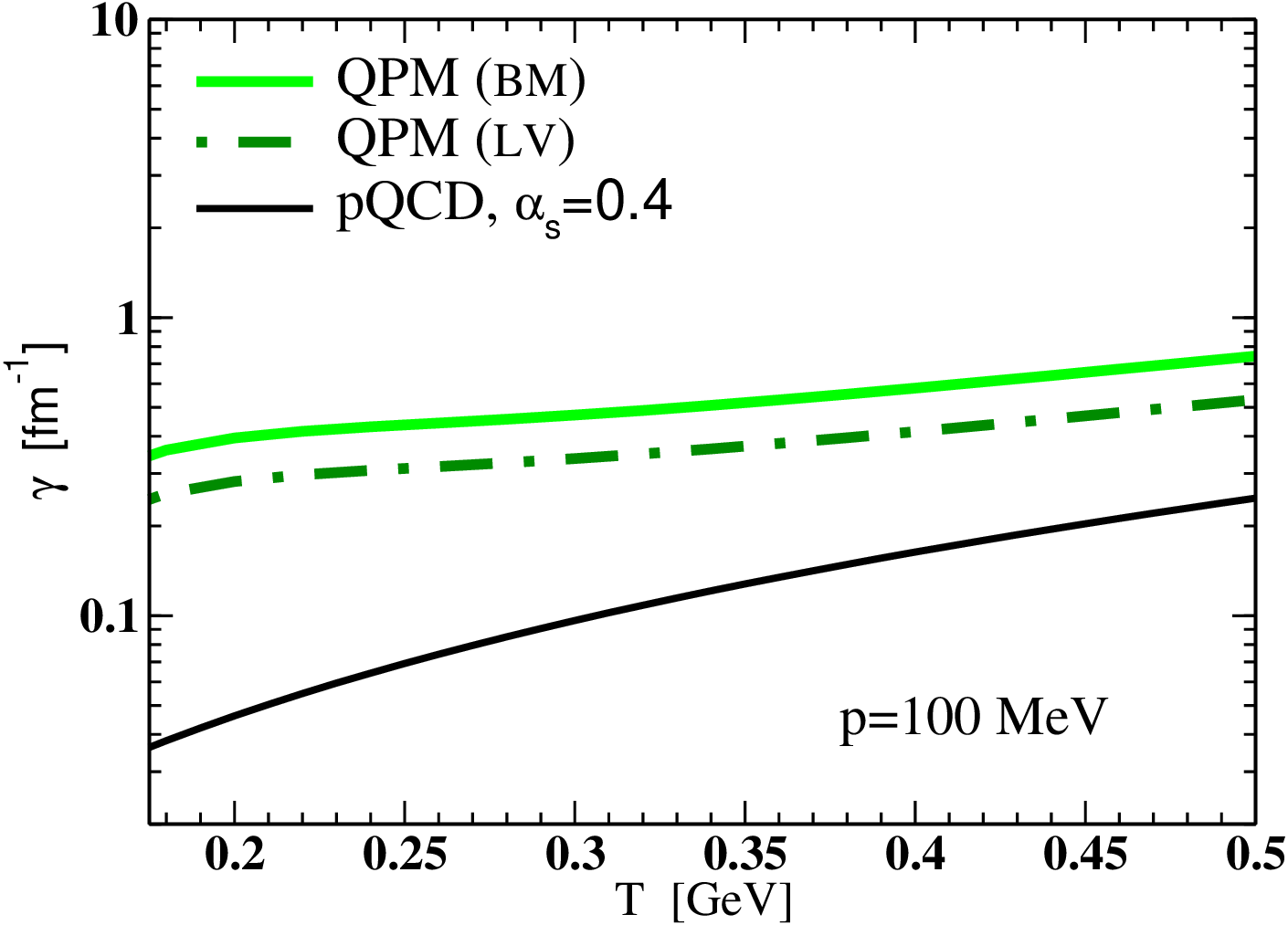}\hspace{2pc}
\caption{Drag coefficients as a function of temperature obtained within the Boltzmann transport approach and Langevin dynamics to 
describe the same experimental data (shown in Fig~\ref{fig4}).}
\label{fig11}
\end{center}
\end{figure}

In Fig.~\ref{fig11} we show the behavior of the drag coefficient $\gamma$ with temperature in QPM by solid line.
Such a drag is evaluated with the same scattering matrix ${\cal M}_{{(q,g)+Q} \leftrightarrow {(q,g)+Q}}$ 
driving the Boltzmann transport. By dotted line we show the same coefficient that is rescaled to
describe the same $R_{AA}(p_T)$,  as the one obtained with Botlzmann dynamics but with  the Langevin dynamics, see Fig. \ref{fig4}
and the related discussion. 
In fact as first observed in ref~\cite{fs}, we need 
a smaller drag coefficient in Langevin dynamics to describe nearly the same experimental results than the Boltzmann transport approach.
In the same figure we have also included for comparison a standard LO-pQCD calculation with a constant coupling. 
This allows us to have an indication of the enhancement for the drag coefficient w.r.t. the LO pQCD to describe 
the experimental data. Moreover we notice that QPM has a weaker temperature dependence of the drag coefficient which is one of 
the key ingredient for a simultaneous description of heavy quark $R_{AA}(p_T)$ and $v_2$.
This has been discussed in Ref.~\cite{Das:2015ana}, but within a Langevin and not with a Botlzmann approach 
and not including the impact of coalescence, see Section IV.

The physics behind the different temperature dependence of QPM w.r.t. pQCD is in the increase of the non perturbative
dynamics as the temperature decrease that in a QPM is induced  by the fit to the lQCD thermodynamics. 
The former implies a coupling $g(T)$ which increase as $T \rightarrow T_c$ in a way that nearly compesates
the decrease of the density resulting in a quite weak temperature dependence of the drag coefficient.

For a constant coupling and massless particles the drag $\gamma$ would go like $1/T^2$,
solid black line in Fig.\ref{fig11} with the strong decrease
mainly driven by the decrease of the bulk density scatterers that is proportional to $1/T^3$.
The same $T$ dependence appears in AdS/CFT because also in such a case the strength of the interaction,
the coupling to the medium, is not temperature dependence. Of course, however, in such a case
the absolute value is much larger of about one order of magnitude with respect to the pQCD one, see also Fig. \ref{fig13}.

\section{Hadronization for charm quarks via coalescence and fragmentation}
Hadronization dynamics plays an important role in determining the final spectra and therefore the $R_{AA}(p_T)$ and
$v_2(p_T)$ in both the light and heavy quark sector \cite{Greco:2003xt,Greco:2003vf,Greco:2003mm,rappv2}.
In particular for heavy quarks it is generally expected that a coalescence mechanism is in action especially
at low and intermediate $p_T$.  We consider here a hybrid model of coalescence plus fragmentation discussing
in detail its impact on both $R_{AA}$ and $v_2$.

In our approach the hadronization hypersurface is determined by the isothermal surface of the bulk dynamics, which
means that is determined stopping collisions between the light particles  and the heavy quarks
when the temperature of a cell drops below the critical temperature that has been fixed to T=155 MeV.

The contribution to hadronization due to coalescence is evaluated according to:
\begin{eqnarray}
\label{eq-coal}
\frac{d^{2}N_{M}}{dP_{T}^{2}}&=& g_{M} \int \prod^{2}_{i=1} \frac{d^{3}p_{i}}{(2\pi)^{3}E_{i}} p_{i} \cdot d\sigma_{i}  \; f_{q_i}(x_{i}, p_{i})\nonumber \\ 
&\times& f_{M}(x_{1}-x_{2}, p_{1}-p_{2})\, \nonumber\\
&\times& \delta^{(2)} \left(P_{T}- p_{T,1} - p_{T,2} \right)
\label{eq_coal}
\end{eqnarray}
where $d\sigma_{i}$ denotes an element of a space-like hypersurface, $g_{H}$ is the 
statistical factor to form a colorless hadron from quark and antiquark with spin 1/2. $f_{q_i}$ are the 
quark (anti-quark) distribution in phase space. $f_{M}$ is the Wigner function and describes the spatial 
and momentum distribution of quarks in the $D$ meson. 

In the Greco-Ko-Levai (GKL) approach~\cite{Greco:2003vf}  for a heavy meson the Wigner function 
is taken as a Gaussian of radius $\Delta_{x}$ in the coordinate  and $\Delta_{p}$ in the 
momentum space,  these two parameters are related  by the uncertainty principle $\Delta_{x}\Delta_{p}=1$,  
\begin{eqnarray} 
f_{M}(x_{1}, x_{2}; p_{1}, p_{2}) =\, 8
\exp(x_r^2/2\Delta_x^2) \,  \exp(p_r^2/2\Delta_p^2)
\end{eqnarray}
where the relative coordinates $x_{r}=x_{1} - x_{2}$ and $p_{r}=p_1-p_2$ are the quadri-vectors for the relative coordinates.
A pattern confirmed by all the groups, despite differences in the details, is that an hadronization by coalescence
is dominant at low momenta \cite{gossiauxv2,he1,Song:2015ykw,Cao:2016gvr}, since the early work in Ref.s \cite{Greco:2003vf,rappv2}.
We determine the width parameter $\Delta_p$ by requiring that the mean square charge radius
of $D^+$ meson is $<r^2>_{ch}=0.43 \, \rm fm$ according to quark model. Given that for our wave function:

\be
<r^2>_{ch}=\frac{3}{2} \frac{Q_1 m_2^2+Q_2 m_1^2}{(m_1+m_2)^2} \frac{1}{\Delta_p^2}
\ee
with $Q_1=+2/3$ and $Q_2=+1/3$ we find $\Delta_p=0.283\, \rm GeV$. We also include the $D^*$
resonant states suppressed according to the statistical thermal weight with respect to the ground state.

We compute the coalescence probability for each charm quark in the phase space point $(\vec x, \tau, \vec p)$
and then assign a probability of fragmentation as $P_{frag}(\vec x, \tau, \vec p)=1 - P_{coal}(\vec x, \tau, \vec p) $.
Therefore the charm distribution function undergoing fragmentation is  evaluated convoluting the momentum of heavy quarks 
which do not undergone to coalescence with the Peterson fragmentation function~\cite{Pet}:
\be
f(z) \propto 
\frac{1}{\lbrack z \lbrack 1- \frac{1}{z}- \frac{\epsilon_c}{1-z} \rbrack^2 \rbrack}
\label{fg}
\ee
where $z=p_D/p_c$ is the momentum fraction of the heavy meson fragmented from the heavy quark and
$\epsilon_c$ is a free parameter to fix the shape of the fragmentation function.
As discussed in the next Section the $\epsilon_c$ parameter will be determined assuring that the available
data on D meson production in $pp$ collisions are well described by a fragmentation hadronization mechanism.
Finally, given the momentum distribution of the charm quarks 
obtained solving the Boltzmann equation the momentum distribution of D meson is calculated summing up
the $D$ meson spectrum obtained via coalescence with the one from  fragmentation.

\section{Comparison to the experimental observables }

\begin{figure}[ht]
\begin{center}
\includegraphics[width=18pc,clip=true]{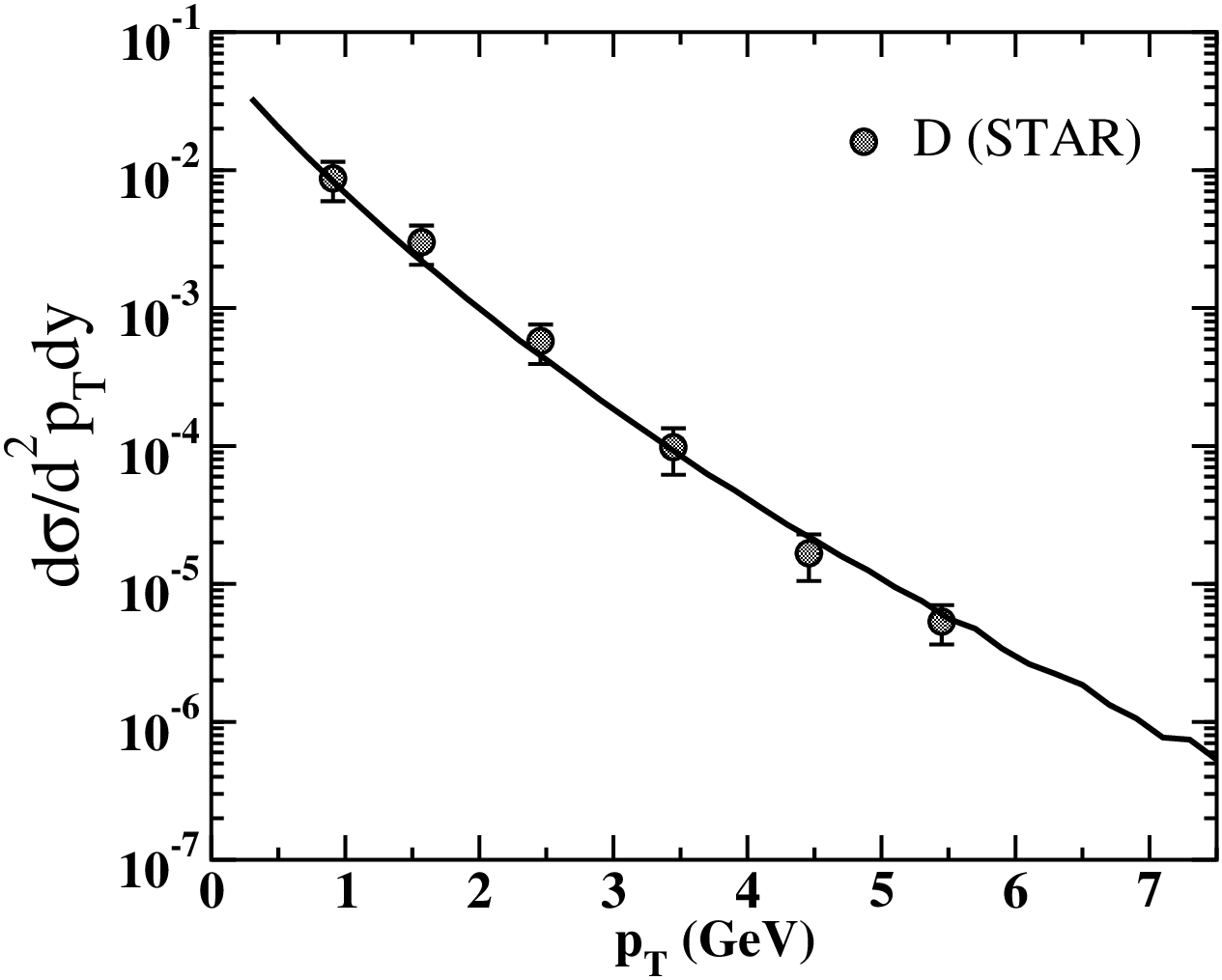}\hspace{2pc}
\caption{The $p_T$ distribution of D mesons, obtained from the fragmentation of charm quarks in p+p collisions, are
compared with the experimental data from the STAR Collaboration,  taken from Ref.~\cite{pprhic}.}
\label{fig1}
\end{center}
\end{figure}

\begin{figure}[ht]
\begin{center}
\includegraphics[width=18pc,clip=true]{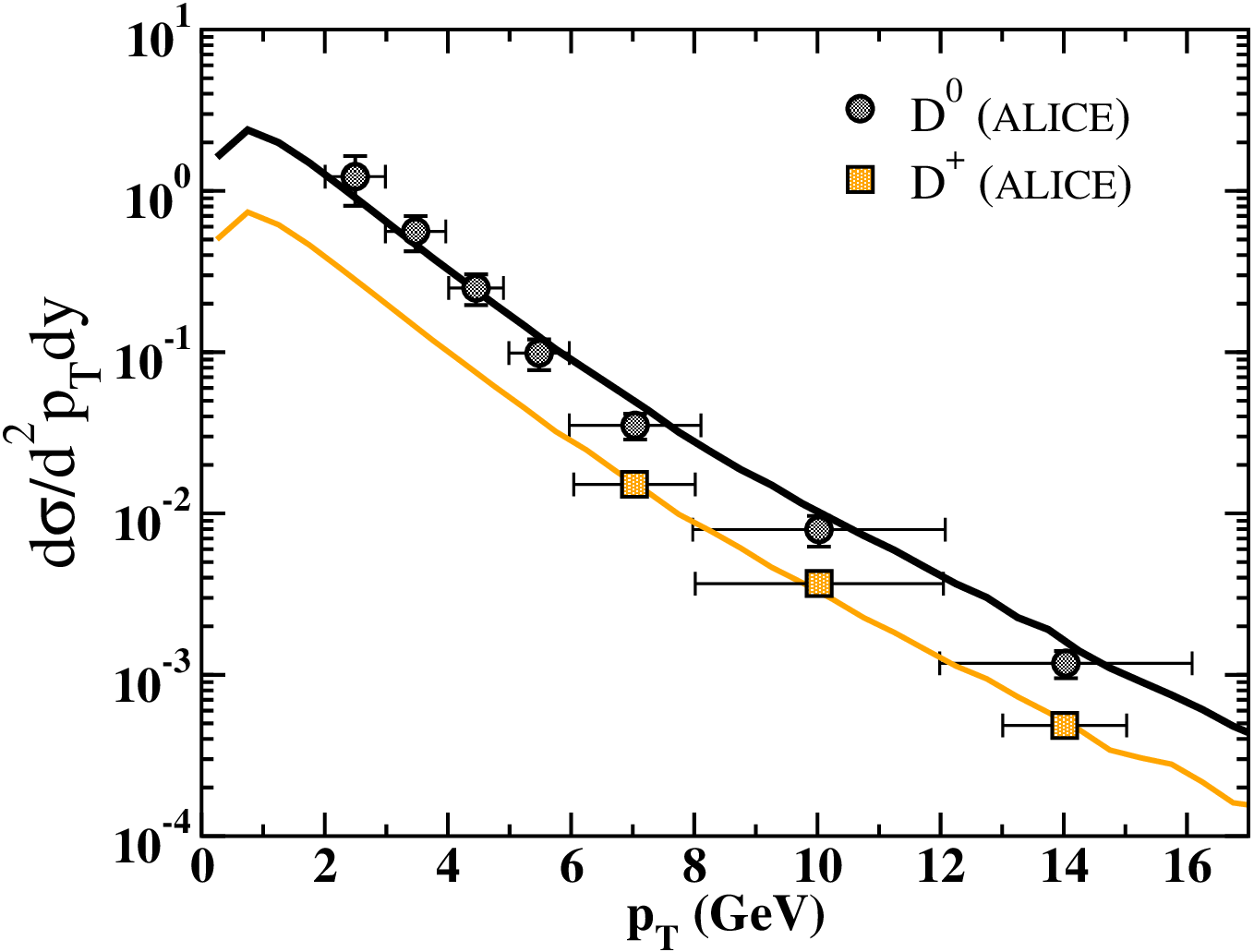}\hspace{2pc}
\caption{The $p_T$ distribution of $D^0$ and $D^+$ mesons, obtained from the fragmentation of charm quarks in p+p collisions, are
compared with the experimental data from the ALICE Collaboration,taken from Ref.~\cite{pplhc}. The experimental points are 
an extrapolation from 7 TeV to 2.76 TeV.}
\label{fig2}
\end{center}
\end{figure}

We present in this section the comparison of the results we get for the nuclear modification factor $R_{AA}$ and  
for the elliptic flow $v_2$ with the experimental data .
We calculate the nuclear suppression factor, $R_{AA}$,  as the ratio of our initial heavy meson distribution at $t=\tau_i$ 
and final heavy meson distribution at $t=\tau_f$ as:
\be
R_{AA}^{c,D}(p_T)=\frac{f_{c,D}(p_T,\tau_f)}{f_{c,D}(p_T,\tau_i)} .
\label{eq.RAA}
\ee
where $f_{c,D}(p_T,\tau_f)$ indicates the momentum distribution already integrated in the $r-$space and in the rapidity
range $|y_z|\leq 0.5$ for charm or $D$ mesons. The $f_{c,D}(p_T,\tau_i)$ is the same distribution we employ for $pp$ collisions,
and that is shown in Fig. \ref{fig1} and \ref{fig2}, see discussion below.
However we note that at LHC where the shadowing effect is expected to be large  \cite{Eskola:2009uj} ,we have also
considered the case where in AA we start from an initial distribution function that is not the $f_{c}(p_T,\tau_i)$ that goes in the
denominator of Eq.\ref{eq.RAA} but is given by:
\be
f^{SW}_c(p_T,\tau_i) = f_c(p_T,\tau_i) * S(p_T)
\ee
where the shadowing function $S(p_T)$ is a parametrization of EPS09 \cite{Eskola:2009uj,ALICE:2012ab},
already integrated in the pertinent rapidity region and over the $r-$space.

In $p+p$ collisions, we convoluted the charm quarks distribution according to the 
Fixed Order + Next-to-Leading Log (FONLL) calculations, taken from Ref.~\cite{Cacciari:2005rk, Cacciari:2012ny}
with the Peterson fragmentation function~\cite{Pet}
to obtain the D meson spectra. As mentioned the free parameter $\epsilon_c $ in the fragmentation function, in Eq. ~\ref{fg},
has been fixed by comparison to the $D^0$ meson production in $p+p$ collisions at RHIC energy as measured
by STAR \cite{pprhic}. With  $\epsilon_c$=0.006 we obtain the spectrum shown in Fig.\ref{fig1} by solid line.. 
In Fig. \ref{fig2} we show $D^0$ and $D^+$ meson spectra in $p+p$ collisions at LHC energy by solid black
and light red lines, obtained with $\epsilon_c$=0.02,
and compare to the experimental results \cite{pplhc} that is an extrapolation from 7 TeV to 2.76 TeV.
With this initial conditions for charm distribution fucntion and their fragmentation function
we proceed to evaluate the D mesons spectra in heavy-ion collisions at RHIC and LHC. 

For $Au+Au$ collisions at $\sqrt{s}= 200$ AGeV,
the initial conditions for the bulk in the $r$-space are given by the standard Glauber condition, while in the $p$-space
we use a Boltzmann-Juttner distribution function up to a transverse momentum $p_T=2$ GeV and
at larger momenta mini-jet distributions as calculated by pQCD at NLO order \cite{Greco:2003xt}. 
The initial maximum temperature at the center of the fireball is $T_0=345$ MeV and
the initial time for the simulations is $\tau_0=0.6$ fm/c, as commonly assumed in hyrdrodynamical simulation
\cite{Romatschke:2007mq,Heinz,Song:2011hk,Niemi:2011ix}, which is about corresponding to the $\tau_0 \cdot T_0 \sim 1$ criterium. 
In our calculation quarks and gluons are massive in order to reproduce the lattice QCD equation of state,
as mentioned in Section II.
In the $p$-space the charm quarks are distributed according to the 
Fixed Order + Next-to-Leading Log (FONLL) calculations, taken from Ref.~\cite{Cacciari:2005rk, Cacciari:2012ny}. 
In the coordinate space HQ are distributed according to number of binary nucleon-nucleon collisions ($N_{coll}$).
corresponding to a constant cross section $\sigma_{NN}=40 \, mb$ at RHIC and  a$\sigma_{NN}=72\, mb$ at LHC.

The dynamical evolution of the bulk is constrained by an $\eta/s=1/4\pi$,  as 
discussed in section II, in such way that the model reproduces the experimental data on 
the bulk spectra and elliptic flow \cite{Ruggieri:2013ova,Plumari:2015cfa}. 
When the system reaches locally the critical temperature
the one body distribution functions of heavy quark are frozen and used to get  the 
momentum distribution. This allows to evaluate the nuclear modification factor and the elliptic flow,  of the D mesons 
by means of the hadronization model described in the previous section.

\begin{figure}[ht]
\begin{center}
\includegraphics[width=18pc,clip=true]{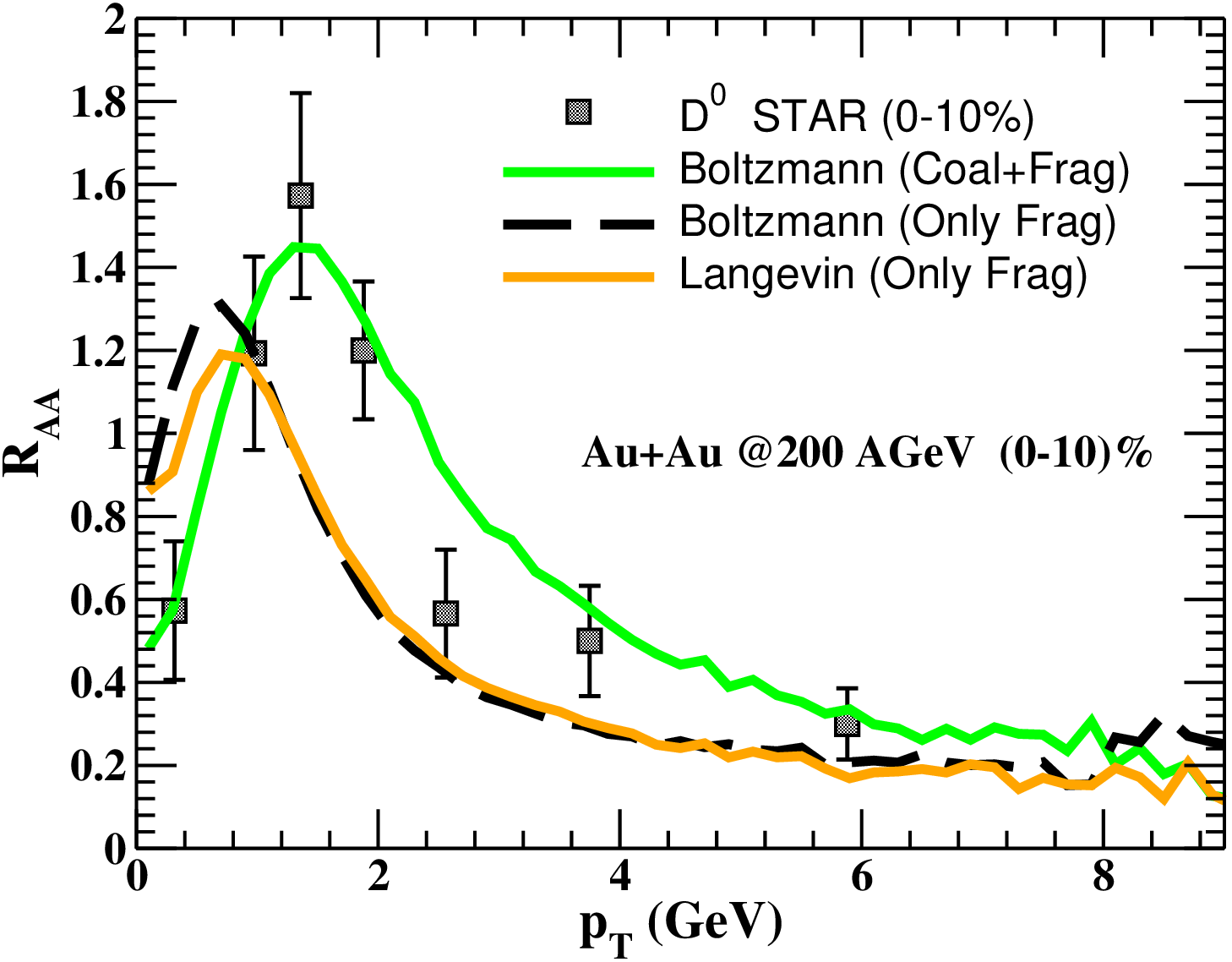}\hspace{2pc}
\caption{D meson $R_{AA}$ in $Au+Au$ collisions at $\sqrt{s}=200\,\rm AGeV$  and centrality $0-10\%$ compared to STAR data. 
Experimental data has been taken from Ref~\cite{rhicraa}.}
\label{fig4}
\end{center}
\end{figure}

In Fig.~\ref{fig4}, the $R_{AA}(p_T)$ as a function of $p_T$ in $Au+Au$ collisions at $\sqrt{s}=200 \,\rm AGeV$  
for  centralities $0-10\%$  that we obtained within our model calculation is depicted and compared with
the experimental data measured at RHIC energy~\cite{rhicraa}. 
In this figure we indicate the impact of coalescence on $R_{AA}$ showing 
the $R_{AA}$  we obtain considering only fragmentation (dashed line) along with the results obtained 
including the coalescence  mechanism plus fragmentation (green solid line). 
We observe that the coalescence  implies an increasing of the $R_{AA}$ for momenta larger 
than $1 \,\rm GeV$ thus a reduction of the suppression.
This is due to the hadronization mechanism which implies that a D mesons from
coalescence of one light quark and a charm quark get a momentum kick with respect to the D mesons obtained 
from fragmentation that on the contrary has a reduced momentum w.r.t.
the original quark according to $\epsilon_c$ in Eq.\ref{fg}. This along with the fact that charm spectrum decreases  with $p_T$
implies that the final spectrum of D meson does not scale with the spectrum of the original charm. An 
increasing in the number of particle in the region of $p_T>1 \,GeV$ is observed. 
At larger momenta, see Fig.\ref{fig7}, fragmentation becomes anyway the dominant mechanism
of hadronization. Such a decrease of coalecence impact that appears naturally in our model seems
to be necessary to describe the $p_T$ depedence observed experimentally.
It has to be mentioned here that the trend of the experimental data at low $p_T$ supports also the coalescence 
as the mechanism of heavy quark hadronization. Heavy quark hadronization only fragmentation could
not describe the marked low $p_T$ bump of the experimental data.

\begin{figure}[ht]
\begin{center}
\includegraphics[width=18pc,clip=true]{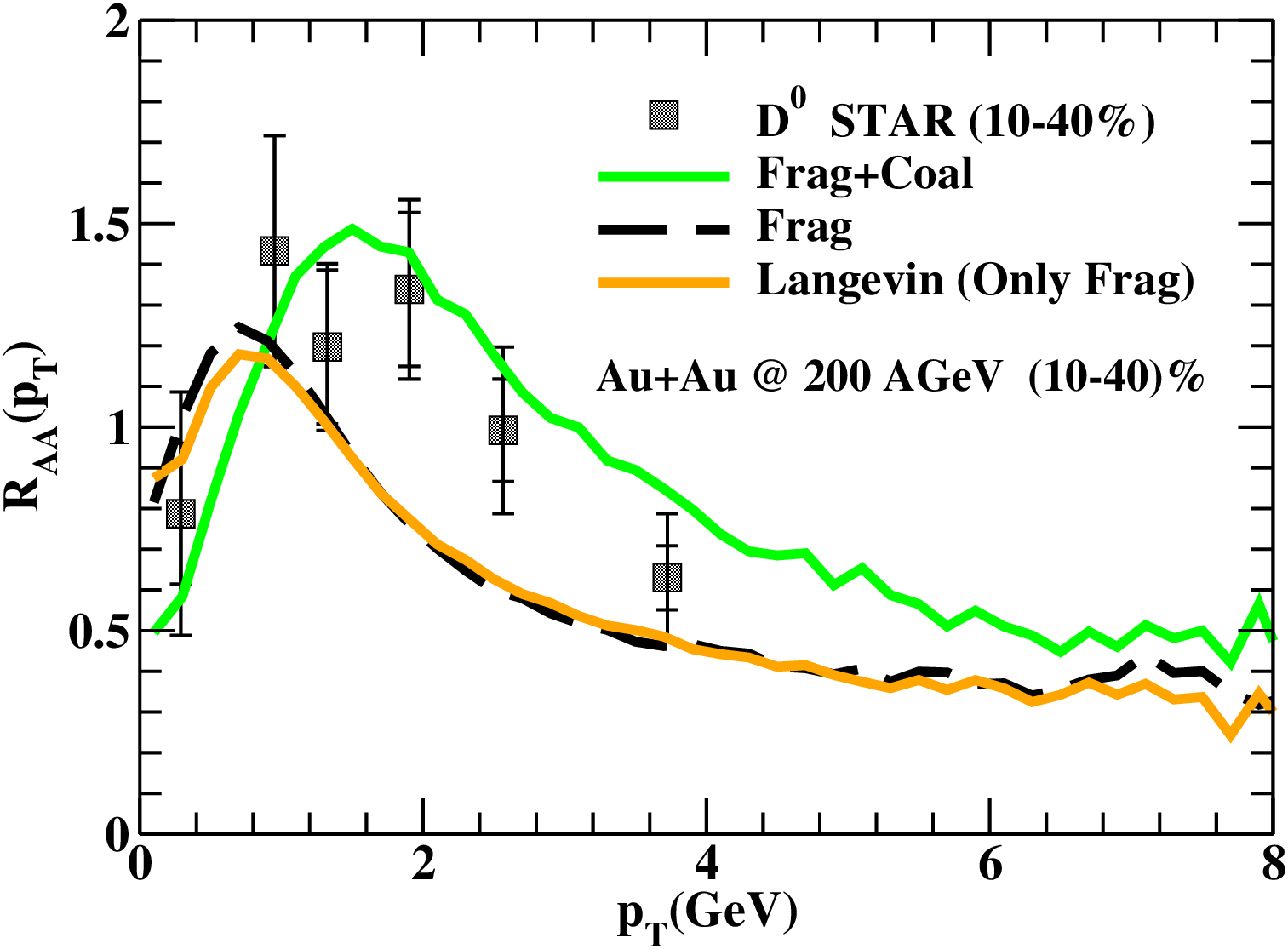}\hspace{2pc}
\caption{D meson $R_{AA}$ in $Au+Au$ collisions at $\sqrt{s}=200 \,\rm AGeV$  and centrality $10-40\%$ compared to STAR data. 
Experimental data has been taken from Ref~\cite{rhicraa}.}
\label{fig5}
\end{center}
\end{figure}

It is shown in ref.~\cite {fs} that a non negligible difference arises between the Langevin and the Boltzmann approach to 
describes the HQ momentum evolution in QGP. In the present study at RHIC we evaluate the $R_{AA}(p_T)$ in $Au+Au$ 
collisions at $\sqrt{s}=200 \,\rm AGeV$  within the Langevin dynamics for the centralities $0-10\%$ and compare the results 
obtained within the Boltzmann transport approach. For the details of the Langevin simulations of HQ dynamics in QGP, 
we refer to Ref.~\cite{fs,Das:2015ana,Das:2016cwd}. In Fig.~\ref{fig4} we show the variation of $R_{AA}(p_T)$ obtained within the 
Langevin dynamics using only fragmentation as the hadronization mechanism and compare the results obtained within the 
Boltzmann transport approach. As shown in Fig.~\ref{fig4}, we can obtain a very similar  $R_{AA}(p_T)$ within both the 
Langevin dynamics as well using Boltzmann transport approach. However, the drag coefficient needed to predict a similar
$R_{AA}(p_T)$ within both the approach has to be rescaled down by about $30\%$ as shown in Fig.\ref{fig11}.
On the other hand it is noteworth that once the drag coefficient is rescaled by a constant (momentum independent factor)
the prediction for $R_{AA}(p_T)$ is nearly identical in quite a large range of $p_T$.
An additional comment is however necessary. As discussed in Ref.\cite{fs} for some ideal case, the comparison of a Langevin
dynamics with a Boltzmann one depends also on the way fluctuation-dissipation theorem (FDT) is implemented.
We report here the case in which Langevin and Boltzmann are more similar. This occurs when the
drag $\gamma(T,p_T)$ is evaluated from the scattering matrix while the diffusion coefficient are determined 
as $B_L=B_T=TE\gamma$. As known there are other possible choice which would lead in general to larger differences
w.r.t. the Boltzmann dynamics, see \cite{fs}. 

Using the same interaction, as of Fig.~\ref{fig4}, within Boltzmann transport approach, we proceed to compare 
the results at RHIC for a different centrality class as well as at LHC colliding energy.
In Fig.~\ref{fig5}, we shown the $R_{AA}$ as a function of $p_T$ in $Au+Au$ collisions at $\sqrt{s}=200 \,\rm AGeV$  
for  centralities $10-40\%$  and compared with the experimental data measured at RHIC energy~\cite{rhicraa}. 
By black dashed line the results obtain within only fragmentation and the green solid line obtained with 
fragmentation plus coalescence. In this centrality also we are getting 
reasonable agreement with the experimental data again, once the coalescence is included along with the fragmentation as the 
hadronization mechanism.

\begin{figure}[ht]
\begin{center}
\includegraphics[width=18pc,clip=true]{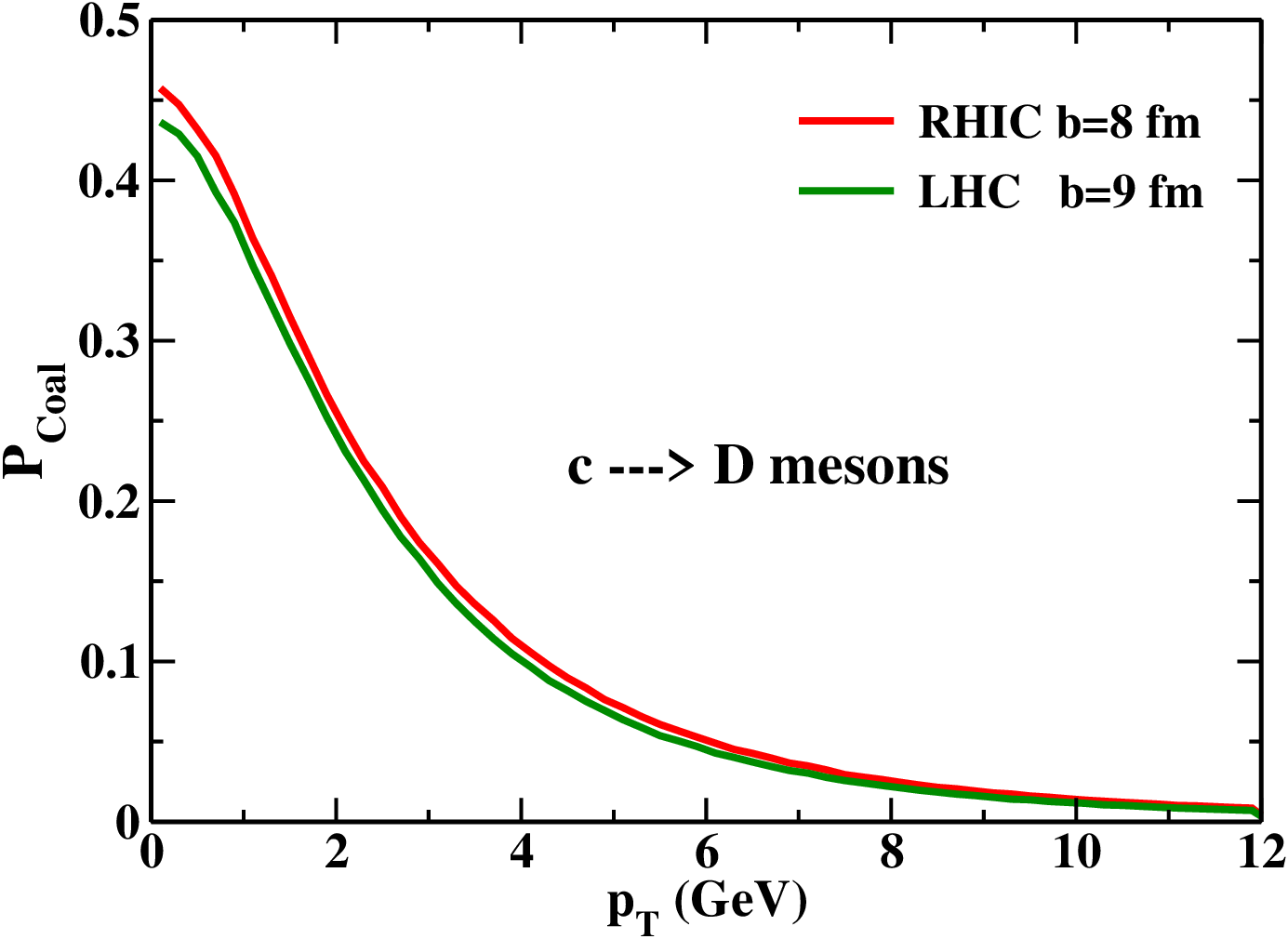}\hspace{2pc}
\caption{Coalescence probability for a charm going to one of the mesons ($D^+, D^0, D^{*0}, D^{+*}$)
as a function of transverse momentum at RHIC and LHC.}
\label{fig7}
\end{center}
\end{figure}

In  Fig.~\ref{fig7} the probability of a charm hadronizing to a $D$ mesons ($D^0,D^+,D^{*0},D^{+*}$) 
through coalescence is depicted as a function of $p_T$.
The charm quark hadronization probability to all hadrons, i.e. including charm baryons,  is set to one in the $p\rightarrow0$ limit,
as usually done by several groups \cite{he1,bass,Song:2015ykw,Nahrgang:2016lst}. The hadronization 
probability decreases with momentum as the coalescence probability involves the product of two distribution 
functions that are decreasing with $p_T$. 
We found at LHC energy the coalescence probability is only marginally smaller than RHIC due to the harder
charm quark distribution at LHC than RHIC.

\begin{figure}[ht]
\begin{center}
\includegraphics[width=18pc,clip=true]{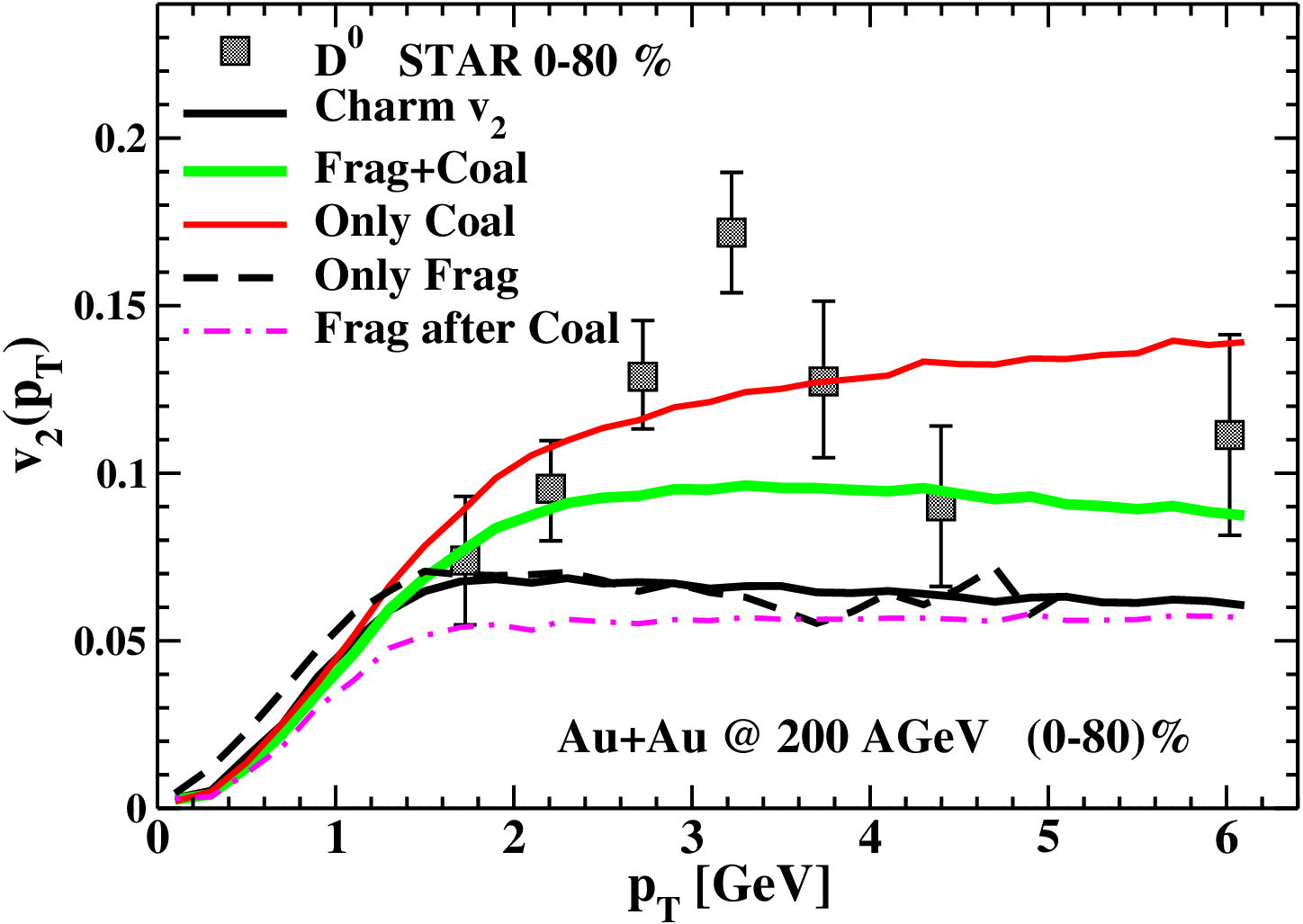}\hspace{2pc}
\caption{D meson elliptic flow in $Au+Au$ collisions at $\sqrt{s}=200 \,\rm AGeV$  and centrality $0-80\%$ compared to STAR data.
Experimental data has been taken from Ref~\cite{rhicv2}.}
\label{fig6}
\end{center}
\end{figure}

In Fig.\ref{fig6} are depicted the results for the elliptic flow as a function of momentum in $Au+Au$ collisions 
at $\sqrt{s}=200 \,\rm AGeV$  for $b=8\, fm$ that on average corresponds to the centrality  $0-80\%$.
We show explicitely the different contributions allowing a direct access to the role played by initial charm
$v_2$ and by coalescence and fragmentation.
The black line indicates the elliptic flow we get for the charm quark, obtained within the Boltzmann transport approach, 
without considering any hadronization mechanism, while the dashed black line indicates the $v_2$ for D mesons that we 
obtain considering only the fragmentation as hadronization mechanism.
We observe that the $v_2$ is  similar in the two cases with a little shift in the low momentum for the D meson case.  
This is because the fragmentation implies that the D-mesons $v_2$ at a given transverse momentum 
is the result of the fragmentation of a charm quark $v_2$ with a slightly larger transverse momentum. 

\begin{figure}[ht]
\begin{center}
\includegraphics[width=18pc,clip=true]{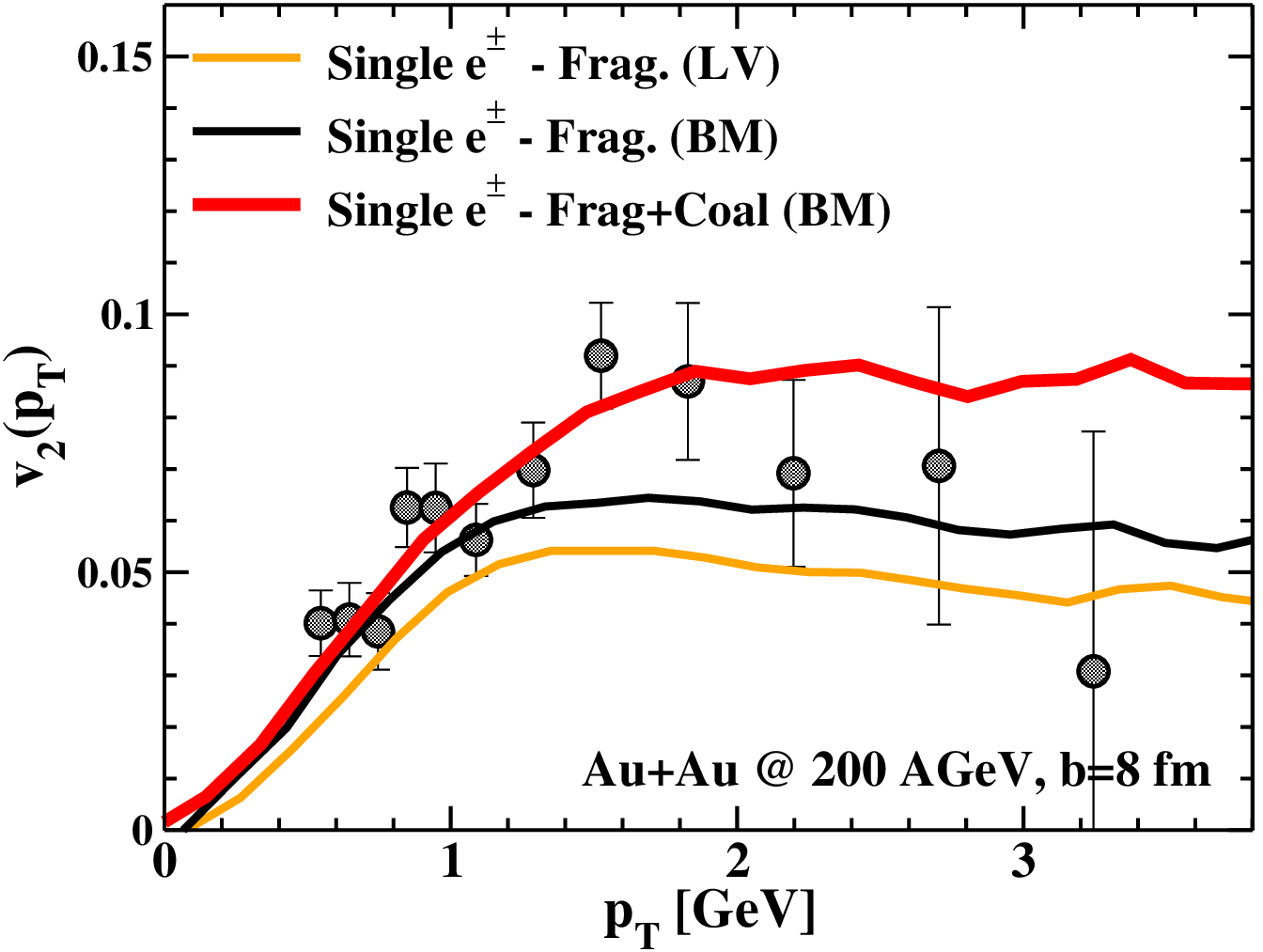}\hspace{2pc}
\caption{Single electron elliptic flow in $Au+Au$ collisions at $\sqrt{s}=200 \,\rm AGeV$  in minimum bias compared to PHENIX data.
Experimental data has been taken from Ref~\cite{phenixelat}.}
\label{fig15}
\end{center}
\end{figure}
If coalescence plus fragmentation mechanism is included for the hadronization, the $v_2$ of the D-mesons increases with respect 
to the elliptic flow of charm quarks by about a $30\%$, solid green line
in Fig. \ref{fig6}. This is because the D meson is the result of the coalescence of a charm 
quark and a light quark and thus the D mesons anisotropy in  momentum space reflect both the heavy quark 
and light quark anisotropies in momentum space. 
The solid red line shows the $v_2$ of  D mesons produced only via coalescence. As expected, the 
$v_2$ developed only coalescence is larger than the $v_2$ developed due to coalescence plus fragmentation.
 It can even lead to an increase of about a factor of two 
at $p_T>2 \, \rm GeV$. This is due mainly due to the large $v_2$ of light quarks w.r.t. charm.
The solid magenta line indicates the elliptic flow produced by the fragmentation of charm once also coalescence
has been switched on. In this case the elliptic flow is smaller with respect 
to that obtained when fragmentation is the only hadronization mechanism, indicated by 
dashed double dotted line. 
This last result is an indirect consequence of the phase space selection 
implicit in the coalescence mechanism that favors the quark pairs that are more correlated
and hence those having momenta closer to the collective flow direction have a large coalescence
probability. The ensemble of charm quarks left over from the coalescence process have a smaller
$v_2(p_T)$ than the one of all the charm quark before hadronization.
This is the reason why the $v_2$ is small for D mesons 
fragmented after coalescence than the D meson formed when only fragmentation is considered.

\begin{figure}[ht]
\begin{center}
\includegraphics[width=18pc,clip=true]{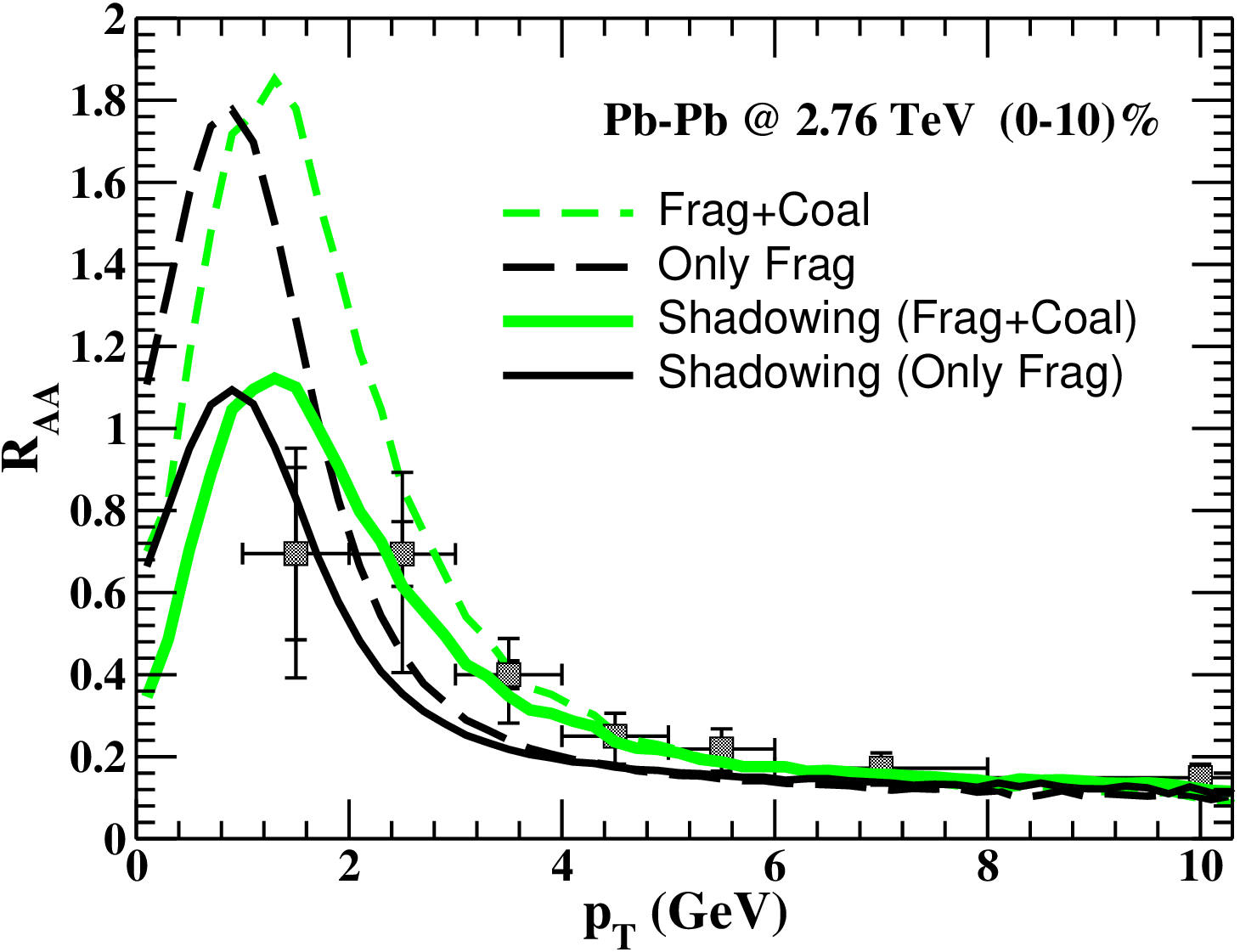}\hspace{2pc}
\caption{D meson $R_{AA}$ in $Pb+Pb$ collisions at $\sqrt{s}=2.76 \,\rm ATev$  and centrality $0-20\%$ compared to ALICE data.
Experimental data has been taken from Ref~\cite{aliceraa010}.}
\label{fig8}
\end{center}
\end{figure}

We have seen in Fig.\ref{fig4} and \ref{fig5} that rescaling the interaction by about a $30\%$ Langevin
gives results very similar to a Boltzmann dynamics. We extend here the comparison to elliptic flow
showing in Fig.\ref{fig15} a comparison with the $v_2(p_T)$ obtained in Langevin simulations.
Langevin dynamics generates about a $15\%$ smaller elliptic flow at charm level which
propagates also to the elliptic flow of single electrons from D mesons decay, shown in Fig.\ref{fig8}
by orange solid line along with the one of the Boltzmann case in black solid line.
We mention that if for the Langevin case the interaction is not scaled down by $30\%$ then the elliptic flow
would be quite similar to the Boltzmann case, but the corresponding $R_{AA}(p_T)$ would be quite
smaller. We also show in Fig.\ref{fig15} by solid thick red line our prediction for the single $e^\pm$ with the
Boltzmann dynamics and hadronization by coalescence plus fragmentation that indeed is in good agreement
with the experimental data from PHENIX. We just remind experimental single electrons come also from
B meson decay and are expected to be significant  for $p_T> \, 2 \,\rm GeV$.

Using the same QPM drag coefficient as at RHIC, we have carried out a simulation of  $Pb+Pb$ collisions at $\sqrt{s}= 2.76$ ATeV 
for centralities $0-10\%$ and $30-50\%$. In this case the initial maximum temperature in the center of the fireball 
is $T_0=490$ MeV and the initial time for the simulations is $\tau_0\sim 1/T_0 =0.3$ fm/c.
In Fig.~\ref{fig8} the results for the $R_{AA}$ at  $0-10\%$ centrality are depicted with only fragmentation and 
fragmentation plus coalescence. In LHC, as of RHIC,  we observe that the coalescence  implies an 
increasing of the $R_{AA}$ for momenta larger than $1 GeV$. 
As evident, the effect of coalescence is less significant at LHC energy 
than at RHIC energy. This is because the effect of coalescence depends on the slope of the charm 
quark momentum distribution. For a harder charm quark distribution the gain in momentum reflects in a smaller increase
of the slope spectrum, instead if the charm quark distribution decreases faster in momentum then the same momentum gain due to coalescence
will result in a stronger increase of the spectrum. For a harder charm quark distribution like at LHC energy 
the impact of coalescence is therefore less pronounced, despite still we see that it leads to a better agreement with the
experimental data also at LHC.

\begin{figure}[ht]
\begin{center}
\includegraphics[width=18pc,clip=true]{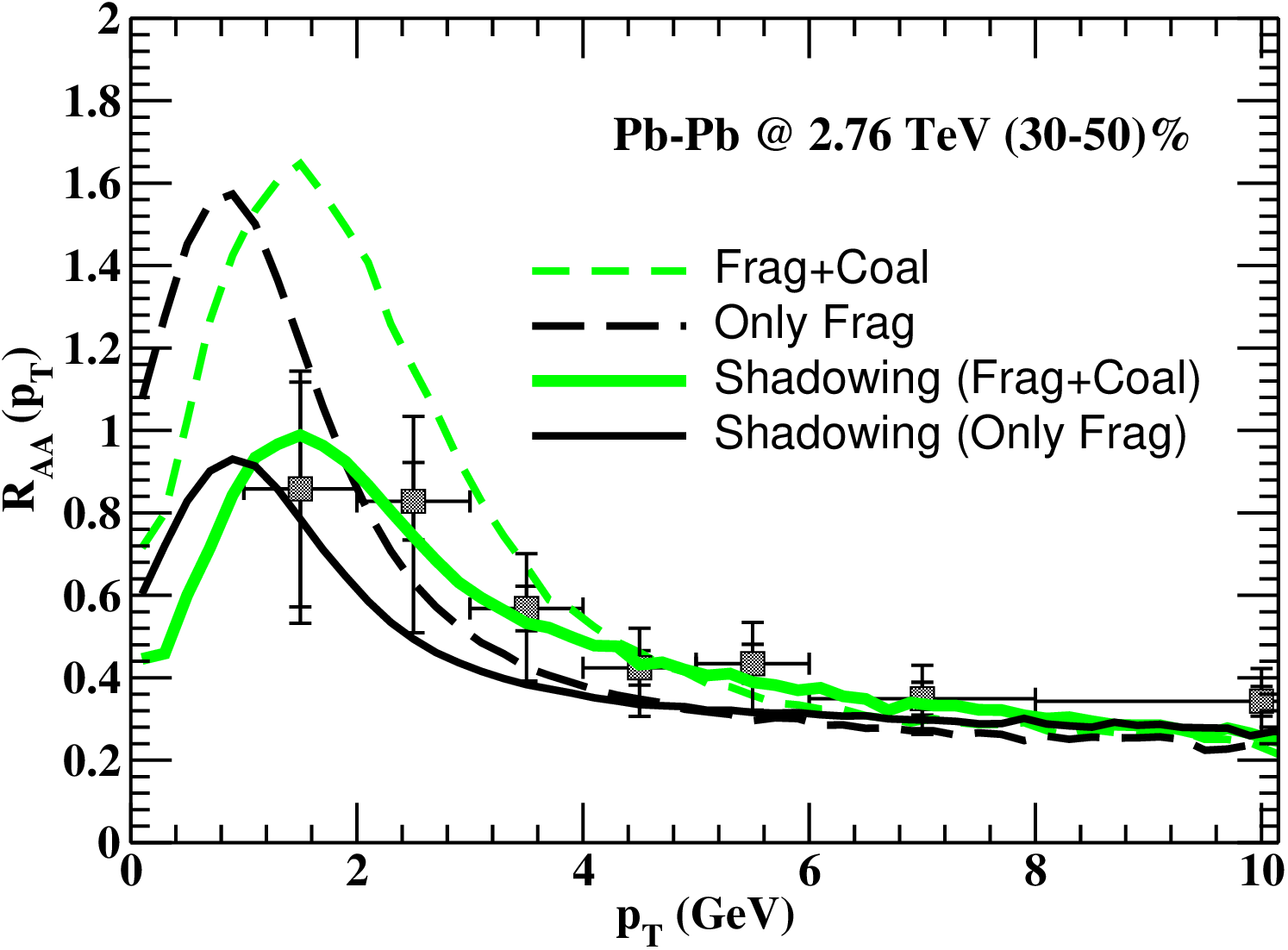}\hspace{2pc}
\caption{D meson $R_{AA}$ in $Pb+Pb$ collisions at $\sqrt{s}=2.76 \, \rm ATev$  and centrality $30-50\%$ compared to ALICE data.
Experimental data has been taken from Ref~\cite{aliceraa010}.}
\label{fig9}
\end{center}
\end{figure}

\begin{figure}[ht]
\begin{center}
\includegraphics[width=18pc,clip=true]{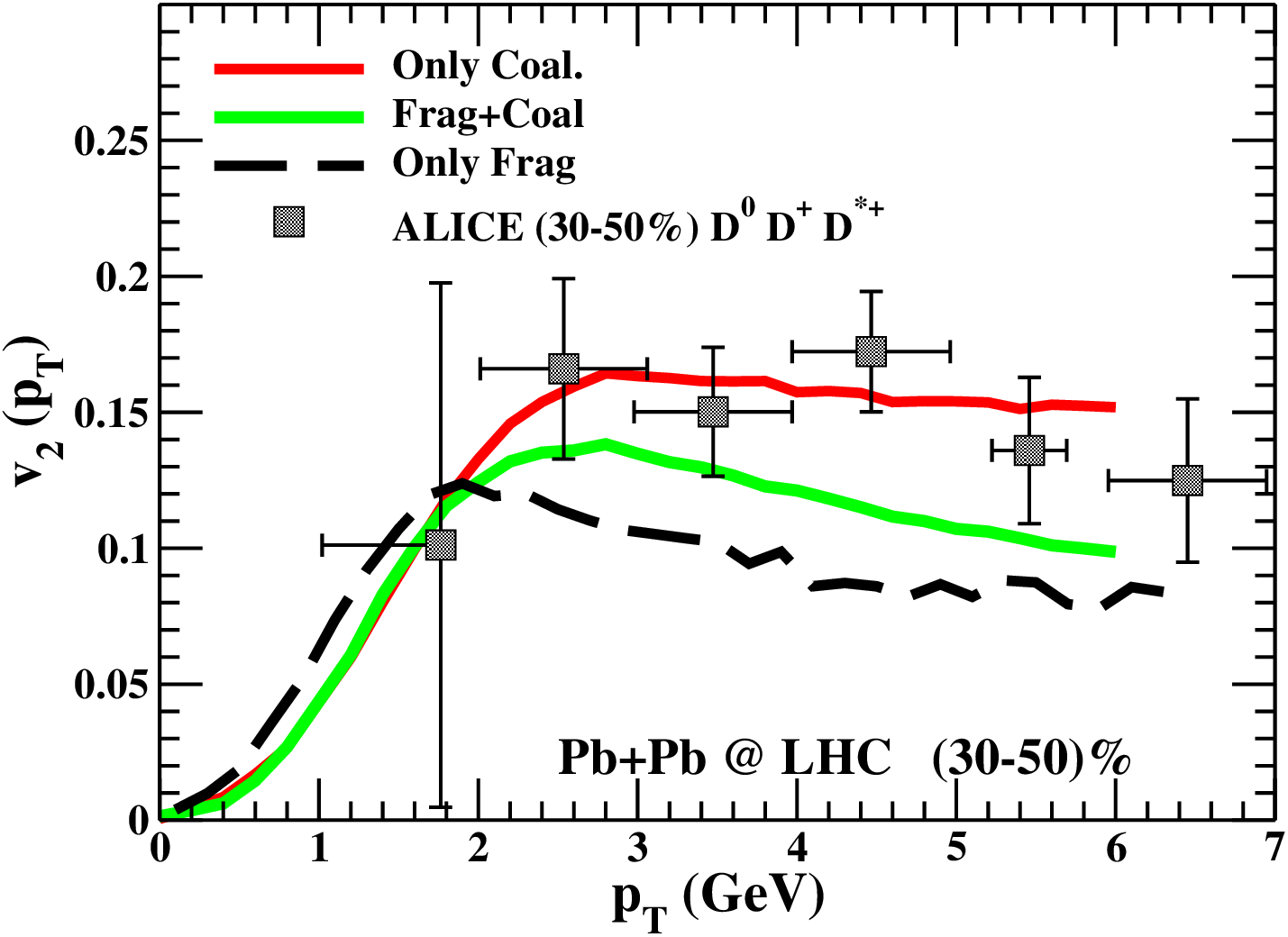}\hspace{2pc}
\caption{D meson elliptic flow in $Pb+Pb$ collisions at $\sqrt{s}=2.76 \, \rm ATeV$  and centrality $30-50\%$ compared to ALICE data.
Experimental data has been taken from Ref~\cite{Abelev:2013lca}.}
\label{fig10}
\end{center}
\end{figure}

In Fig.~\ref{fig9} we present $R_{AA}$ with respect to $p_T$ for more peripheral collisions at $30-50\%$.
We see a similar coalescence effect as in centrall collision and also in this case it allows a much better
description of the experimental data when the shadowing is included.
Indeed the data and our calculations seem to clearly show a shadowing effect in agreement with EPS09 \cite{Eskola:2009uj}
within the still large uncertainties at low $p_T$ in the data.
In Fig.~\ref{fig10}  as expected coalescence increasing both the $R_{AA}$ and $v_2$ and bring the results close to the data, towards a 
simultaneous description of heavy meson $R_{AA}$ and $v_2$. At LHC energy also, the $v_2$ is
significantly smaller for D mesons 
fragmented after coalescence than the D meson produced due to only fragmentation. 
We notice the very recent data \cite{Abelev:2013lca} are an average of the measurements in $Pb+Pb$
at  $\sqrt{s}=5.02 \, \rm ATeV$ that we include here because are the only available data on $v_2$ with a not too
larger uncertainty. On the other hand in our modeling the results increasing the beam energy up to
5.02 ATeV would only affect the elliptic flow by few percent which is quite negligible with respect to the present
uncertainties. 

It is important to note that the impact of coalescence cannot be mimicked by heavy quark diffusion,
in fact at variance with it coalescence leads to an enhancement of both $R_{AA}(p_T)$ and $v_2(p_T)$.
At RHIC energy, considering coalescence plus fragmentation as the charm quark hadronization mechanism, 
the D meson get about $30\%$ of $v_2$ from the light partons as a consequence of coalescence. But still the major
part of the D meson $v_2$ (about 65-70$\%$) is coming from the heavy quark diffusion within QGP (heavy quark-bulk interaction). 
On the other at LHC energy the D meson get about 20-25$\%$ $v_2$ from the light partons as a consequence of coalescence 
where as about 75-80$\%$ is coming from the heavy quark diffusion within QGP (heavy quark-bulk interaction).
We remind that the T dependence of the drag coefficient is very important, as pointed out in Ref. \cite{Das:2015ana}, to 
obtained the charm quark flow coming from diffusion. 

\section{Transport Diffusion coefficients and thermalization time}

\begin{figure}[ht]
\begin{center}
\includegraphics[width=19pc,clip=true]{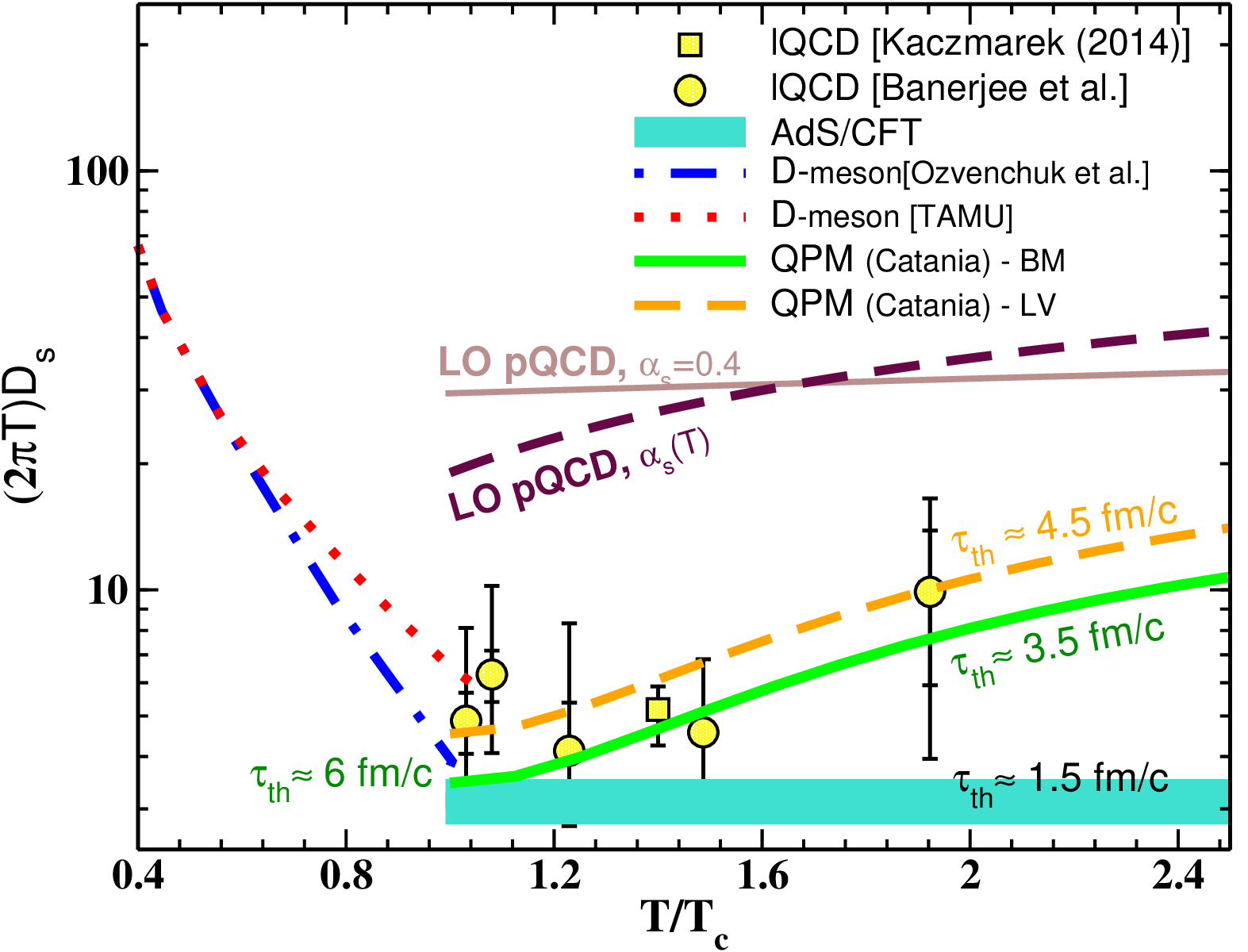}\hspace{2pc}
\caption{Spatial  diffusion coefficient as a function of temperature obtained within the Boltzmann transport approach and Langevin dynamics 
to describe experimental data, along  with the results from lattice QCD. We have also shown the results obtained within other models.
Spatial diffusion coefficient for the QPM model employed to prediction for $R_{AA}$ and $v_2$ within a  Boltzmann (BM)
dynamics (green solid line) and with a Langevin (LV) one (orange dashed  line) \citep{Das:2015ana}; By symbols
quenched lQCD \citep{Banerjee:2011ra} (circles) \cite{Kaczmarek:2014jga} (square); 
model calculations based on LO pQCD \cite{vanHees:2004gq,moore} (solid and dashed brown
lines) and AdS/CFT rescaled to match the energy density of QCD plasma \cite{Gubser:2006qh};
in dotted line is shown the $D_s$ coefficient for D meson in hadronic matter \cite{He:2011yi}.}

\label{fig13}
\end{center}
\end{figure}

The space diffusion coefficient $D_s$ is the most significant transport parameter that quantifies
the interaction of heavy quarks with the medium that is directly related to the thermalization time and can be
evaluated also in lattice QCD, whose more recent calculation are shown by circles and squares in Fig. \ref{fig13}, where
a standard quantification of the space diffusion coefficient is done in terms of the adimensional quantity $2\pi T D_s$. 
In Fig.~\ref{fig13}, we also show T dependence of the spatial diffusion coefficient underlying the
predictions for $R_{AA}(p_T)$ and $v_2(p_T)$ described in the previous section.
The space diffusion obtained with Boltzmann equation is plotted by green solid  line and the one about a $30\%$
larger corresponding to the Langevin simulations. 

For comparison we also show results obtained within LO pQCD for constant coupling, $\alpha_s=0.4$,
by solid brown line which is independent of temperature. The case with a running coupling $\alpha_s(T)$, 
maroon solid line, leads to a weak temperature dependence. 
However it is well known that such a value of the pQCD $D_s$
cannot describe at all the small $R_{AA}(p_T)$ observed as well as the $v_2(p_T)$.
Indeed from a more rigorous theoretical point of view it has been shown that for charm quarks the
perturbative expansion does not show any sign of convergency \cite{CaronHuot:2007gq} (unless one is in the
weak coupling regime $\alpha_s<0.05$), and the value shown is only indicative of a LO term.
Nicely both lattice QCD and the phenomenology discussed here as in all other approaches cited in the introduction
firmly agree with a $D_s$ is much smaller than this LO pQCD estimate.
In the same plot, we have also shown the results for D meson in the hadronic phase  by dotted line the
results of Ref. \cite{He:2011yi}, which are also very similar to \cite{Tolos:2013kva,Das:2011vba} 
and by dash dotted line the one in Ref. \cite{Ozvenchuk:2014rpa}.
The $D_s$ is directly related to the drag coefficient $\gamma$ driving the HQ 
in medium evolution that has been discussed in Section II. We have $D_s=T/M*\gamma(p \rightarrow 0)$ and has to be noticed
that in kinetic theory one, and on a more gneral ground, $\gamma$ is expected to be proportional to $M$, hence
$D_s$ should be about mass independent provinding a general measure of the QCD interaction.
It is interesting to estimate the thermalization time, $ \tau_{th} \equiv \gamma^{-1}(p\rightarrow 0)$, for charm quarks:
\be
\tau_{th} = \frac{M}{2\pi T^2} (2\pi T D_s) \cong 1.8 \, \frac{2\pi T D_s}{(T/T_c)^2} \,\, \rm fm/c
\ee
where we have substituted a charm quark mass $M=1.4 \,\rm GeV$ and we have written $\tau_{th}$ in terms
of a dimensional quantities to facilitate a direct evaluation. For example at $T\simeq 2\,T_c$  the dashed orange lane as
the central value of lQCD gives $2\pi T D_s=10$ which means a $\tau_{th} \simeq 4.5 \,\rm fm/c$ as indicated directly in
the figure. Similarly one can easily derive the values in the other points of the plot, in the QGP branch.
Notice that none of the behavior shown corresponds to a constant thermalization time which would imply a 
drag $\gamma$ completely T-independent, i.e. $2\pi T D_s \propto T^2$. 
In particular the flat $2\pi T D_s$ corresponds to a thermalization time
proportional to $1/T^2$ which is the case of AdS/CFT \cite{Gubser:2006qh} 
and pQCD with constant coupling.

The predictions relative to RAA and v2 obtained by means of the Boltzmann equation with the QPM model 
indicate a thermalization time that stays in the the range $3.5-6 \, \rm fm/c$ which at LHC
energy is smaller then the lifetime of the QGP, especially in central collisions.
However it has been noticed in Ref.\cite{Beraudo:2014boa}  that the $R_{AA}(p_T)$ still significantly
deviates, also at low $p_T$ smaller than 3-4 GeV, from the one expected in case of full thermalization. 
This is an aspect that has to be
investigated in more details to spot the origin of such a deviation. However it has to be considered
that our estimate of thermalization time is done in the $p \rightarrow 0$ limit, should take into account 
for the decreasing of the interaction with the increasing of the momentum which implies a termalization 
time of about a $ 50 \%$ larger or more, once average over momentum is done.

We also mention that the $2\pi TD_s$ estimated in the Langevin case, orange dashed line,
is nearly identical to the one of the $T$-matrix approach \cite{he1}, while the Boltzmann one is quite close
to the one from the PHSD transport approach which is based indeed on a dynamical quasi-particle
model that includes also finite widths \cite{Song:2015ykw}.

It may be mentioned that our $D_s$ is marginally below the lattice QCD data point near $T_c$.  
A final assessment requests in principle the inclusions of initial state fluctuations~\cite{Cao:2014fna,Plumari:2015cfa}, 
which helps to develop a large suppression than the averaged one, but such an effect is in practice quite nominal 
pointing to a somewhat smaller drag coefficient, hence, a larger $D_s$. Such an effect is expected to be within a $10\%$.
Also as shown in Ref. ~\cite{Das:2017dsh, Das:2015aga}, the pre-equilibrium 
phase may affect the heavy quark suppression pointing also to a somewhat larger $D_s$ for a better agreement with
data from the experiments. 
However the impact of these further aspects is in general less relevant with respect to the differences between 
Langevin and Boltzmann approaches and quite smaller than current experimental error bars and the systemtic uncertainties
in lattice QCD calculations. Certainly we are reaching a stage where it will become relevant to include them,
especially with the upcoming experimental data expected with much smaller error bars and the 
new data on the triangular flow $v_3$~\cite{Nahrgang:2014vza,Nahrgang:2016wig, Prado:2016szr} that however is beyond the scope of the present paper.
Certainly our work gives a further contribution in showing that the extraction of the diffusion coefficient from
the phenomenology of heavy-ion collisions is possible and reliable.

\section{Summary and conclusions}
We have studied the heavy quark propagation in QGP at RHIC and LHC within a Boltzmann transport approach.
We start with a charm quark initialization describing the D mesons production in $p+p$ 
collisions reasonable well at both RHIC and LHC energies within an hadronization by indipendent fragmentation . 
The heavy quark and the bulk 
interaction has been taken into account within a QPM which is able to reproduce the lattice QCD equation of
state and their in-medium evolution has been treated with a Boltzmann transport equation with some
comparison to the Langevin one. 
The hadronization of heavy quark in AA collisions is described by means of a hybrid model of fragmentation plus coalescence. 
Using the same interaction we have calculated the $R_{AA}$ and $v_2$ of heavy meson at different centrality class 
as well as different colliding energies. Our model gives a good description for D meson
 $R_{AA}$ and $v_2$ both at RHIC and LHC within the still significantly large systematic and
statistical uncertainties.
There are three main ingredients that we identify as playing a key role toward the agreement to the
experimental data. The first one is the QPM that enhances the heavy quark bulk interaction near $T_c$ with 
respect to a mere decrease of the drag coefficient $\gamma$ with $1/T^2$. 
This non-perturbative behavior catched by a QPM modeling fitted to the lattice QCD thermodynamics,
was identified as the key ingredient for the build-up of 
a large $v_2$~\cite{Das:2015ana} in a Langevin approach and is confirmed within the
Boltzmann approach mainly discussed in the present paper. 

Implementation of heavy quark hadronization by means of a hybrid model of fragmentation plus coalescence 
help to increase both the $R_{AA}$ and $v_2$ close to the data, if the coalescence is regulated to
exhaust hadronization in the zero momentum region, the effect is quite large and essential.
Finally as discussed  in ref.~\cite {fs} for test calculations and ideal cases,
once the interaction is tuned to very similar $R_{AA}(p_T)$
the Boltzmann approach is more efficient in reproducing the elliptic flow w.r.t. Langevin
an effect that for the QPM model is of about a $15\%$.

The underlying $2\pi T D_s$ diffusion coefficient rises about linearly with temperature $T$ 
and leads to an initial thermalization time of about $3\,\rm fm/c$ at the maximum initial temperature
reached at LHC energy ($T \simeq 3 T_c$) increasing only to about a $6\, \rm fm/c$ around $T_c$. This would suggest that the
core of charm quarks produced may be essentialy kinetically thermalized at the time where most of hadornization
occurs.
A finding that was certainly unexpected before starting the endeavor to study heavy quark diffusion
in the quark gluon plasma and that can have important implications for the understanding also of the quarkonia
production.
It may be mentioned here that the hadronic rescattering, while generally not affecting $R_{AA}$, give a further contribution 
to D meson $v_2$ that is in the range of 10-20$\%$ ~\cite{Das:2016llg} depending  on the $T_c$ assumed that is generally in 
the range $155-175$ MeV. Considering the fact that we use $T_c$=155 MeV, our results
will not be affected significantly by the hadronic rescattering. In any case hadronic rescattering would lead to an improvement 
of the agreement between the experimental data and the present modeling.

Our result and the estimated $D_s(T)$ shows nice agreement with lattice QCD data within the
still significantly large uncertainties. This feature is shared nowdays by most of the modelings,
as discussed in  \cite{Greco:2017qm}; in particular our estimate of the value and T dependence
of $D_s(T)$ is quite close to \cite{he1,Song:2015ykw}. Even if it should be quantified if also the agreement
to the data is quantitatevely similar. Other apporaches can lead to a current
estimate that indicates similar values but with a weaker temperature dependence. In general
this can be traced back to some difference in the evolution of the bulk matter and/or to a different
impact of coalescence. However especially for this last point upcoming data on $\Lambda_c/D$
and $D_s/D$ will shed a new light and allow for more stringent constraints on the hadronization 
mechanism.Therefore in the next future it would be important that all the phenomenological approaches
aiming at the evaluation of the heavy transport coefficient provide their predictions also for 
such ratios as a function of transverse momentum.
Furthermore we think that we are reaching a stage where starts, to be appropriate, to have a comparison
to several experimental data as a function of energy and centrality that is statistical quantified.
This effort has already been started by some groups \cite{Xu:2017hgt}, but it
will also be more meaningful and powerful with the new upcoming data that are expected to carry much
smaller error bars and to be extended to lower $p_T$, the region more relevant for the transport coefficient
determination. Given that current uncertainties in the phenomenological approach is comparable
if not smaller than the current lattice QCD data it is desireable that in the following years the latter will be
able to reduce the present systematic uncertainities. 
Certainly Open Heavy Flavor physics in ultra-relativistic heavy-ion collisions
is showing to really have the potential to link the phenomenology to lattice QCD for studying
the transport properties of the Hot QCD matter.

\vspace{2mm}
\section*{Acknowledgments}
We acknowledge the support by the ERC StG under the QGPDyn Grant n. 259684. 
V.G. acknowledge also the warm hospitality at the UCAS in Beijing under the 
President International Fellowship Iniziative (2016VMA063).

\end{document}